\documentclass[english]{article}
\usepackage[T1]{fontenc}
\usepackage[latin9]{inputenc}
\usepackage[includeheadfoot,margin=2cm]{geometry}
\usepackage{fancyhdr}
\usepackage{amsthm}
\usepackage{amsmath}
\usepackage{graphicx}
\usepackage{multirow}
\usepackage{array}

\makeatletter
\numberwithin{equation}{section}
\numberwithin{figure}{section}

\newcommand{\argmin}{\operatornamewithlimits{argmin}}
\newcommand{\argmax}{\operatornamewithlimits{argmax}}

\makeatother

\usepackage{babel}
\begin{document}

\title{\textsc{Matching Mechanisms For Real-Time Computational Resource
Exchange Markets}}

\author{\textsc{Joseph W Robinson, jwrobins@andrew.cmu.edu, carnegie mellon
university}\\
\textsc{Aaron Q Li, aaron@potatos.io, carnegie mellon university}}

\maketitle
\begin{abstract}
In this paper we describe matching mechanisms for a real-time computational
resource exchange market, Chital, that incentivizes participating
clients to perform computation for their peers in exchange for overall
improved performance. The system is designed to discourage dishonest
behavior via a credit system, while simultaneously minimizing the
use of dedicated computing servers and the number of verifications
performed by the administrating servers. We describe the system in
the context of a pre-existing system (under development), Vedalia
\cite{715Project}, for analyzing and visualizing product reviews,
by using machine learning such as topic models. We extend this context
to general computing tasks, describe a list of matching algorithms,
and evaluate their performance in a simulated environment. In addition, 
we design a matching algorithm that optimizes the amount
of time a participant could save compared to computing a task on their
own, and show empirically that this algorithm results in a situation in which it is 
almost always optimal for a user to
join the exchange than do computation alone. Lastly, we use a top-down approach
to derive a theoretically near-optimal matching algorithm under certain distributional
assumptions on query frequency.
\end{abstract}

\section{Introduction}

\paragraph*{\textmd{The concept of decentralized and distributed systems has
materialized in many revolutionary real-world applications: Peer-to-peer
(P2P) downloading for file sharing, Bitcoin (BTC) for virtual-currency,
and many more. However, there is yet a similar system under the context
of sharing computing resources that has achieved a reasonable level of success.
On the other hand, there is an increasing use of computationally intensive
algorithms on personal and Internet data, especially those developed
in machine learning and artificial intelligence research. Typically,
in production systems, these computations are carried out only for
large models in a centralized fashion using large scale cloud computing
platforms or high-performance clusters in data centers owned by large
corporations. But as applications of machine learning and artificial
intelligence become ubiquitous, there is a need for computing smaller
models given a reasonable amount budget using small computing devices,
such as mobile phones in the context of mobile apps, or office / lab
computers in the context of enterprise applications and scientific
research.}}

\paragraph*{\textmd{For such a system to exist, there must be a way for computing
nodes to discover each other and exchange computing tasks. A simple
way to make this work is to create a local computational resource
exchange market that matches buyers, who have tasks to compute, and
sellers, who can offer computational resources. In this paper, we formalize
this problem and devise a list of matching algorithms, a set of evaluation
metrics, and show both empirical and theoretical properties of these
algorithms in our experiment results under the context of computing
and sharing topic models of product reviews.}}

\paragraph*{\textmd{The paper is organized as follows: Section \ref{sec:Background}
introduces the background of Vedalia and Chital systems. Vedalia is
a system for computing topic models of product reviews, of which its
scalability and performance issues motivated the design of a computational
resource exchange system, Chital. Section \ref{sec:Matching-Problem}
formalizes the problem in the general context. Section \ref{sec:Evaluation-Metrics}
introduces the key evaluation metrics for performance. The matching
algorithms are given in Sections \ref{sec:Heuristic-Algorithms}
and \ref{sec:Deriving-Theoretically-Optimal}. Simple heuristic algorithms
are given first in Section \ref{sec:Heuristic-Algorithms}, followed
by an empirically stable algorithm that obtains Nash equilibrium with respect to
 users leaving the system with high likelihood under many conditions. We also derive 
 a distributional-based algorithm for matching and present results of several matchers
 obtained via simulation.}}

\section{Background \label{sec:Background}}

\subsection{Vedalia}

\paragraph*{\textmd{The Vedalia system \cite{715Project} was developed in late
2014 as a review analyzer for Amazon products. The system uses latent
variable models (primarily LDA\cite{OriginalLDA2003}) and natural
language processors to analyze and aggregate review text for display
to the user. The Vedalia system overcome the shortcomings of the current
Amazon system by representing each product by its topic distribution.
Each topic distribution is then displayed to the user via a word cloud
in order to provide multifaceted information about a product in a
way that is intuitive, compact, and aggregate. The system retains
completeness in the sense that the full review set is still available
to the user via an interactive topic-based search feature. }}

\paragraph*{\textmd{In the next stage we intend to bring Vedalia to mobile phones
and expand the system by including review sources from all over the
Internet. However, one challenge we face is the high computation cost
associated with topic modeling. Each modeling process involves thousands
of iterations over a collection of product reviews. Thus, a typical
product with hundreds of reviews requires several seconds of dedicated
computation on a single core. As there are millions of products and
hundreds of millions of product reviews on Amazon alone \cite{amazondata},
maintaining a centralized architecture for computing topic models
is both expensive and inefficient, even using the state-of-the-art
systems such as \cite{LietalWSDM2015,LietalNIPS2013}.}}

\subsection{Chital}

\paragraph*{\textmd{The Chital system was originally designed to address these
challenges by creating a marketplace for topic model computation.
In the original design, the system incentivizes users to share their
phone's computational resource for topic model computation by giving
the users a chance of winning a lottery offered over every time period,
taken from a portion of the advertisement revenue in the Chital
mobile app. }}

\paragraph*{\textmd{In our revised design, we extend its context to general computing
tasks. While we are keeping the option of letting the user opt in
or out of the computation sharing system, we no longer use lottery as
the primary incentive for that purpose. Instead, we focus on devising
matching algorithms such that it is always beneficial for the
users to join the sharing pool rather than computing the results by themselves. }}

\paragraph*{\textmd{In our analysis of matching algorithms we assume
for the worst case: a lottery system does not exist, and a user only
opts-in to computation for the duration of his task.}}

\paragraph*{\textmd{Just like every marketplace and distributed computation
system, the Chital system has to address the issues of bad behavior
and incorrect results. The following four subsections describe our
mechanisms in preventing the user from reporting false computational
result. Malicious users still may, however, potentially gain some
benefits by manipulating the computing capacity measurement (by throttling
CPU usage) to a lower value than what it actually is. Details on an algorithm that 
empirically obtains Nash equilibrium with respect to users leaving the system are given 
in Section \ref{sec:Almost-Nash-Algorithm}.}}

\subsubsection{Marketplace}

\paragraph*{\textmd{A marketplace is constructed in which buyers (those who issue
computing tasks) and sellers (those that rent out their devices' computational
resources to complete task) are matched using a matching mechanism.
Each buyer whose device has sufficient computational power is automatically
entered as a seller for their own query. Then, for each query, we
select two sellers with sufficient computational power to build the
model, where the sum of the two seller's credit is above a predefined
threshold. Let $c_{1}$ and $c_{2}$ denote the credit of the two
sellers, and $p_{1}$ and $p_{2}$ denote the perplexity of the sellers'
results. Then the probability of secondary verification is defined
as: 
\[
1-\frac{1}{3}\Bigg[\frac{1}{1+e^{-(c_{1}+c_{2})}}+2\frac{\min(p_{1},p_{2})}{\max(p_{1},p_{2})}\Bigg]
\]
Here, the first term denotes the sum of the two sellers credits, transformed
to fall in the range (0,1). The second term represents the similarity
of the quality of the two seller's results; thus, results of vastly
different quality have a higher chance of verification. From this
expression, we see that when both sellers' credits are infinity high
and their results have the same perplexity, no secondary verification
is needed. When both sellers credits are relatively low and the perplexity
match is weak, we must verify that the best of the two results is,
in fact, valid. Details on the secondary verification and model evaluations
system are presented in Section \ref{sub:Evaluation-System}. }}

\subsubsection{Credit System}

\paragraph*{\textmd{Credit functions as a 0-sum system in which we begin with
two 0-credit servers for computation. These servers are controlled
by the Chital system and provide a guarantee that at least two compute
nodes will always be available for building topic models.}}

\paragraph*{\textmd{Each user that joins the system begins with 0 credit. When
building a model, the results of the two client (compute) nodes selected
from the marketplace are compared; the model with the greatest log-likelihood
is selected to be returned to the initial requesting user. A single
credit is then transferred from the user that generated the model
with the lowest log-likelihood to the user that generated the model
with the greatest log-likelihood. Assuming that all users are using
our software to generate valid results, this creates a system in which
each user has an expected credit 0 over time. However, a ``bad''
user that sends invalid results in an attempt to get more tickets
will almost always produce the model with the lower log-likelihood,
causing the credit distribution to shift from the bad to good users
and increasing the likelihood that results from this bad user will
go through secondary verification in the future.}}

\subsubsection{Lottery System }

\paragraph*{\textmd{In our original design, lottery money is allocated after
each ad payout period as a fixed proportion of ad revenue and is fed
back into the system to encourage participation in the marketplace.
The specific time period between lottery picks is variable, and depends
on the quantity of ad revenue (this avoids having an hourly lottery
in which a user can at most win a few dollars/cents, a potentially
insignificant quantity). During a lottery pick, a single user is sampled
at random with probability of winning proportionate to the number
of tickets the user has won over the course of the time period. The
full amount of lottery money is then transferred to this selected
user via PayPal.}}

\paragraph*{\textmd{In our new design, the lottery system is no longer an essential
component as it is no longer a necessary incentive for users to participate in
the computation sharing pool. }}

\subsubsection{Evaluation System \label{sub:Evaluation-System}}

\paragraph*{\textmd{The evaluation system evaluates the relative quality between
multiple results submitted from the computing nodes, and makes the
ultimate decision on their acceptance. Assume we have a set of documents
$\{d_{i}=(t_{i1},t_{i2},...,t_{il_{i}})\}_{i=1,...,D}$ and a topic
model $\{\theta_{k}=(\theta_{ik},...,\theta_{Dk}),\,\phi_{k}=(\phi_{k1},\phi_{k2},...,\phi_{kV})\}_{k=1...K}$
(where $D$ is the number of documents, $t_{ij}$ is the token at
position $j$ for document $i$, $K$ is the number of topics, $V$
is the size of vocabulary, $\phi_{kw}$ is the word probability for
topic $k$, and $\theta_{ik}$ is the probability of $i$-th document
possessing topic $k$..}}

\subparagraph*{\textmd{There are three types of evaluations: }}

\paragraph*{Validation\textmd{: Verify that $|1-\sum_{w}\phi_{kw}|<\epsilon$
for all $k=1,...,K$, and $|1-\sum_{k}\theta_{ik}|<\epsilon$ for
all $i=1,...,D$, where $\epsilon$ is the error-tolerance parameter
for the system. If any submitted topic model fails validation, reject
the model. }}

\paragraph*{Perplexity\textmd{: The perplexity is calculated by 
\begin{eqnarray*}
\pi(\{d_{i}\}) & = & \exp(-(\sum_{i=1}^{D}|d_{i}|)^{-1}\sum_{i=1}^{D}\log p(d_{i}|\theta,\phi))\protect\\
p(d_{i}|\theta,\phi) & = & \prod_{j=1}^{|d_{i}|}\sum_{k=1}^{K}\phi_{kt_{j}}\theta_{ik}
\end{eqnarray*}
}}

\paragraph*{\textmd{In other words, the perplexity is the transformed inverse
geometric mean of log-likelihood over tokens given the model. The
perplexity can be evaluated efficiently within sublinear time to the
total number of tokens (since $\phi_{k}$ and $\theta_{k}$ are sparse
vectors with smoothers). The golden-standard of the perplexity is
unknown unless we independently compute a topic model. However, given
two topic model submissions, we can quickly determine which model
is better by evaluating the perplexity on the original set of documents.}}

\paragraph*{Verification\textmd{: If the credit system determines the submissions
are dubious, a verification is performed. The evaluation system picks
a valid model with lower perplexity, run an efficient version of Gibbs
sampling \cite{LietalANU2012,LietalKDD2014,MALLET}) for a few iterations
on a server-side computation cluster, and record the perplexity after
each iteration is completed. If the perplexity deviate from the original
model by a substantial amount within these few iterations, it would
be evident that the original model is not converged, thus it should
be rejected by the system. }}

\section{Matching Problem \label{sec:Matching-Problem}}

\paragraph*{\textmd{The core of the computation marketplace is the design of
a matching mechanism. In this market, a user submitting computing
tasks to the system is called a buyer. A user offering computational
resource to complete tasks submitted by other people is called a seller.
A user can be simultaneously be a buyer and a seller at the same time.
Naturally, each user has different computing amount of computing power
to offer. In reality the amount of computing power a user can offer
may change over time, but only for a small amount. To make this problem
tractable, we ignore the this and assume there is a way to reliably
instantly measure a user's computing power at negligible cost.
However, in addition to that, a user may control or manipulate this
value at any point of time. We discuss this situation in detail in
later sections and propose mechanisms to prevent this.}}

\paragraph*{\textmd{In the context of Vedalia, each computing task submitted
by the buyer must be completed by two sellers independently for verification
purposes. In the context of general computing task,it is often desirable
to make runs of the same computations for the purpose to avoid local
minima (e.g k-mean algorithm, EM algorithm, Gibbs sampling) in addition
to verifications. Consequently, in our discussions of matching problem
we assume each computing task is run twice in total, regardless whether
it is run by two different sellers, or by the buyer himself.}}

\paragraph*{\textmd{The computational marketplace matching problem has many unique
characteristics that are not present in the traditional settings of
matching problems. Namely:}}
\begin{itemize}
\item Both types of vertices (compute nodes and users) arrive online with
respect to some probability distribution, whereas traditional bipartite
matching algorithms assume there exist one group of offline vertices
and another group arrives online.
\item After a computing node finishes its assigned task, the node may become
available again for matching. In traditional matching problems, once
a vertex is matched (or exceeds its capacity/budget as in some variations
of the matching problem, see \cite{Mehta2013}) it cannot be matched
again.
\item There are tradeoffs between strategyproofness and optimizing different
metrics (see Section \ref{sec:Evaluation-Metrics}). Traditional matching
algorithms usually aim to maximize the sum of weights across all matched
edges. 
\end{itemize}

\paragraph*{\textmd{Even without these differences, it is known that finding
a solution for a general graph matching problem is hard if one aims
for a competitive ratio more than 0.5 (\cite{KohtRegev2003}), the
baseline using the greedy algorithm. Some recent work such as \cite{YajunWangChuwaiWong2013}
has shown it is possible to achieve a competitive ratio of 0.526 for
the fractional matching problem (that a vertex can be divided to multiple
``fractions'' so as to match to multiple vertices). In our matching
problem, there are some features and structures we can use to outperform
the greedy algorithm. For instance:}}
\begin{itemize}
\item A computing node has edges connected to all users whose computing
task can be computed by this node within reasonable amount of time
\item Each task submitter is also a computing node 
\item The distribution of computing power and the arriving time of computing
nodes are known.
\end{itemize}

\paragraph*{\textmd{Formally, the problem is defined as the following: Given
a set of computing devices $P=\{p\}$, we define the matching problem
on sequence of graphs $\{G_{t}\}=\{(S_{t},B_{t},E_{t})\}$, where
$S_{t}\subset P$ and $B_{t}\subset P$ are devices acting as computing
nodes (``sellers'') and users (``buyers'') respectively that currently
are in the marketplace at time $t$. Each buyer $b\in B_{t}$ corresponds
to a query $q\in Q_{t}$, and it is possible that multiple buyers
may have the same query. A match $m=(q,(p_{1},p_{2})):(p_{1},p_{2})\in S_{t},q\in Q_{t}$
between two sellers and a query initiates the computation task contained
in the query $q$ to be carried out by devices $p_{1},p_{2}$. When
a device is performing computation for query $q$, it cannot perform
computation for any other queries until the computation is complete.
When both $p_{1}$ and $p_{2}$ finish computations for query $q$,
we call this query completed.}}

\paragraph*{\textmd{When an existing query $q$ is completed
by sellers $(p_{1},p_{2})$ between time interval $[prev(t),t]$,
the buyers $Buyers(q)$ in $B_{prev(t)}$ are no longer present in $B_{t}$,
and the sellers $(p_{1},p_{2})$ are no longer present in $S_{t}$
unless $p_{1}$ corresponds to a query that has not been completed
at timestamp $t$. New queries arrive in the following fashion: whenever
a query $q$ is completed time $t$, $Buyer(q)$ initiates a new query
at time $t+\lambda$ where $\lambda$ is drawn from some distribution
$ $$\pi(\theta_{1})$ with parameters $\theta_{1}$. }}

\paragraph*{\textmd{Naturally, this corresponds to a continuous time model. Devices
share their computing resources when they make a query (submit a task),
and stop sharing when they finish computing their assigned task and
have their query completed, whichever comes later. Our goal is to
find a matching $M_{t}=\{m\}$ over all times $t$ so to minimize some
value under some constraints (see \ref{sec:Evaluation-Metrics}). 
Here the edges $(b,s)\in E_{t}$
contains all the pairs $(b,s)$ such that the computing node's computational
resource is powerful enough to match with the other user's computing
need. }}

\paragraph*{\textmd{To summarize, we have following quantities (timestamp $t$
is dropped for simplicity):}}
\begin{itemize}
\item $p\in P$: Computing devices that may join the system as sellers and
buyers
\item $b\in B$: Current active buyers waiting on query results
\item $s\in S$: Current active sellers providing computation resources
\item $q\in Q$: A query carrying a computation task submitted by a buyer
\item $m\in M$: A matching between a query $q$ and a seller tuple $(p_{1},p_{2})$
\item $\lambda\sim\pi(\theta_{1})$: $ $For each device, the amount of
time until the next query is made after the device's current query is completed
\end{itemize}

\paragraph*{\textmd{In addition, we define following quantities:}}
\begin{itemize}
\item $Buyer(q)$: The device that initiated query $q$
\item $G(q,p_{1},p_{2})$: The gain for $Buyer(q)$ when query $q$ is matched
with sellers $p_{1},p_{2}$
\item $w(q,p_{1},p_{2},k_{1},k_{2})$: The total wait time when $q$ is
matched with sellers $p_{1},p_{2}$, and this matching at $k_{1}$-th
position waiting for seller $p_{1}$, and $k_{2}$-th position waiting
for seller $p_{2}$. The total wait time includes both the amount
of time waiting for a matching to establish and the amount of time
during which actual computation takes place.
\item $\phi(p,q)$: The amount of computation time for seller $p$ to compute
query $q$
\item $\psi(p)$: Time until device $p$ finishes his current computation.
If $p$ is not currently doing any computation, $\psi(p)=0$.
\item $Perf(p)\sim\sigma(\theta_{2})$: The computation performance (measured
in tokens/second) of a device $p$ is drawn from a distribution $\sigma(\theta_{2})$
with parameters $\theta_{2}$
\item $Tok(q)\sim\omega(\theta_{3})$: The size of the computing task associated with 
query $q$ (measured
in number of tokens) drawn from a distribution
$\omega(\theta_{3})$ with parameters $\theta_{3}$.
\item $\mu$: The number of servers. A server is defined as a computing device
that is much faster than others, and which never submits tasks (thus
never generates queries). The inclusion of a small number of servers is helpful in improving
the initial performance of the system when there are very few sellers.
Servers are also helpful for handling queries that have been waiting in the queue for 
too long.
\end{itemize}

\paragraph*{\textmd{The problem of optimizing matches over all timestamps is intractable
without making an unreasonable number of assumptions. Consequently, in our algorithm
designs we often opt to minimize a quantity for a single time duration
or instantaneous timestamp. }}

\section{Evaluation Metrics \label{sec:Evaluation-Metrics}}

\paragraph*{\textmd{There is no single value that measures the absolute performance
of the system. As a result, optimizing for one value often involves tradeoffs
with other objectives. In practice,
we believe the following metrics are the most relevant: }}
\begin{itemize}
\item Average wait time: the average amount of time for fulfilling each
query. The objective function is $\sum_{q}w(q;\cdot)$ where the other
variables of $w(\cdot)$ are observed. This metric measures the overall
throughput of our system. A lower average wait time means our system
is efficient in maximizing the average rate at which queries are fulfilled and results 
returned to the user.
\item Maximum wait time: the maximum amount of time taken to complete any
query, i.e: $\max_{q}w(q;\cdot)$. This metric measures the worst
case wait time for any participating user. A lower maximum wait time gives
a guaranteed time bound for any query and often means that computational resources
are more evenly shared across devices.
\item Slow query ratio: the percent of queries that are fulfilled slower
than if they were computed by the buyers themselves. This metric measures
the overall incentives for the users to use the system as opposed
to doing computation alone. Lower slow query ratio is desirable
in the sense that it means users are more likely to save computing
time by participating in the system. In the event that slow query ration is less than $\frac{1}{2}$ 
for every user, we have that for any user submitting a query the most likely scenario is one in 
which the query is fulfilled faster than the user could compute the result on his own. 
This can also be used to identify the existence of a Nash Equilibrium in which all users 
benefit from staying in the system.
\item Average amount of time saved / wasted: The average amount of time
saved / wasted across all queries for each device. Similar to slow
query ratio, these two measurements give deeper insight into how much
users gain by participating the system.
\item Net gain / net loss per device, and the number of devices with net
gain / loss: These are additional metrics for measuring incentives
to join the system and the stability of the system. In the case of number of devices with 
net loss, a value of 0 means that no users received negative average utility for being 
a part of the system.
\end{itemize}

\section{Heuristic Algorithms \label{sec:Heuristic-Algorithms}}
\subsection{Nash Equilibrium and Strategyproofness}
In designing our heuristics, we pay special attention to two desirable properties of any matching system and optimization metric -- Nash equilibrium and strategyproofness. In the following discussion we let players represent devices in the system, where each player's goal is to minimize the expected amount of time he waits for his queries to be fulfilled.\\
\\
Suppose each player in the system has two options: stay in the network and allow the system to handle fulfillment of his queries, or leave the network and self-fulfill all queries. In a state of Nash Equilibrium, we have that the expected utility of remaining in the system is at least the expected utility of leaving the system. From our player goals, we have that for each player $p^*$ the following should hold:
\begin{gather*}
2 \sum_{q_{p^*}} \phi(p^*, q_{p^*}) \geq \sum_{q_{p^*}} w(q_{p^*}, p_{match}^{(i)}, p_{match}^{(j)}, k_i, k_j)
\end{gather*}
Here, each $q_{p^*}$ represents a query made by $p^*$, $p_{match}^{(i)}$ and $p_{match}^{(j)}$ represents matchings made by the system for a query, and $k_i$ and $k_j$ represent their respective orderings. Thus, the expected amount of time it takes the matching system to compute a result for a player should be no more than the amount of time it would take $p^*$ to perform the computations himself. As such, in expectation it should be beneficial for each player to remain a part of the system in order to prevent a scenario in which the fastest devices continuously leave the system, causing a cascading effect.\\
\\
We define the notion of strategyproofness as such: a matcher is strategyproof if no player can benefit from querying unnecessary products or misreporting his performance. The first of these follows directly from the fact that we require a player to wait until his last query is completed to make the next query. Thus, in expectation submitting unnecessary queries will lead to an increase in wait time and decreased utility of the system for that player.\\
\\
For the second, we must address two options. In the first, a player attempts to overreport his performance by completing computations faster than should be possible given his device hardware. From our verification system, a player who behaves in this way will in expectation have decreased credit, leading to the eventual removal from the system when the player is identified as misbehaving. On the other hand, it should be the case that a player cannot benefit by underreporting his performance. In this situation, a player takes longer than is necessary to perform computations in order to improve his wait times. In order to avoid this, several heuristics are examined that give priority to faster devices. In this case, it is optimal for a player to report the highest possible performance in order to increase his query priority; overreporting will lead to the loss of credit, and thus the optimal strategy is to report one's true performance.

\subsection{Immediate vs Scheduled Matchers}

We examined two primary types of matcher: immediate and scheduled. In an immediate matcher, a matching event is triggered each time a query arrives or a computation completes. With an immediate matcher, we have the capability of providing the fastest possible time-to-match, however we may miss more optimal matchings that could be obtained by waiting for new sellers to make queries and enter the marketplace.\\
\\
This issue is explored using scheduled matchers, which instead perform matchings at a fixed rate without regard for any events occurring within the marketplace. By performing matchings at a fixed rate, it is sometimes possible to obtain a better assignment of devices to queries by waiting for a small amount of time and accumulating a larger pool of sellers. From this, it is often possible to compute a better matching than would have been obtained by an immediate matcher. For example, consider a matcher that simply matches each incoming query, in order of arrival time, with the two fastest sellers available. Suppose that we have two idle servers $p_1$ and $p_2$ such that $Perf(p_1) = Perf(p_2) >> Perf(p_i)$ for all user devices $p_i$. For simplicity, assume all user devices have the same performance. If queries arrive in order from largest to smallest, our matcher will have to wait for the largest tasks to finish before starting the next smallest task, and so on. This results in substantial wait times that can be remedied by performing scheduled matchings, reordering events by task size, and fulfilling queries in ascending order.

\subsection{Query Processing Order}

The order in which queries are processed in one of the key areas of improvement in designing heuristics. We wish to reorder and fulfill queries in such a way that optimize for each of our evaluation metrics. Below, we briefly describe a few heuristics along with the goals of each.

\subsubsection{FIFO}

In the FIFO ordering, a matcher attempts to fulfill queries in the order in which they arrive. In this way, the FIFO ordering defines a truly greedy ordering with respect to arrival time. In a real-time system, this allows matchings to be made extremely quickly. In addition, by ignoring the task size and performance of each seller when ordering queries we obtain a framework whereby a device cannot misreport his performance in order to improve the priority of his tasks. Unfortunately, the FIFO ordering suffers from many pitfalls. As a key example, this ordering makes the system directly dependent on the ordering in which queries arrive, even within a small timeframe. As such, small variations in query arrival times can result in extremely high variance of wait times.

\subsubsection{Minimum Variance}

Each arriving query begins with a priority equal to its number of tokens. Then, in each matching round, we order all unfilled queries in descending order of priority and attempt to fulfill queries in order. After each round of matching, the priorities of all remaining unmatched queries are doubled. Thus, we attempt to reduce in the variance in the number of rounds a query remains unmatched while still giving priority to larger, harder to complete tasks. 

\subsubsection{Hardest to Fulfill}

In the hardest-to-fulfill heuristic, we order all unfilled queries in ascending order of hardness to fulfill, defined as the amount of time it would take the querying device to do the computation alone:
\begin{gather*}
\text{hardness to fulfill} = 2\phi(Buyer(q), q)
\end{gather*}
In this way, we give priority to queries that are hardest to compute faster than the querying device could have done the computation. In this way, we hope to optimize the percent of queries that are returned faster than a device could obtained alone, thus moving towards a system in which having all devices as part of the system is a Nash equilibrium.
\subsection{Seller Selection}

Once a query is selected as the next-in-line to attempt to fulfill, we have two options. We can match the query with one or more available sellers, or we can defer the matching of the query until the next round. For each of the following, we have the option of each performing partial matchings --  a single seller can begin fulfilling a query before the second matched seller is available -- or enforcing complete matchings -- both seller devices must be available for computation before beginning to fulfill a query. We also test probabilistic versions of several selectors, where each query has a random probability of being skipped during each matching round. By adding this random component, the goal is to maintain a distribution of idle sellers that is more robust to sudden influxes of large tasks, resulting in better overall performance.

\subsubsection{FIFO}

In the FIFO heuristic, we match each incoming with the first two sellers in our list. Not surprisingly, this is the worst of our heuristics by most metrics. While a FIFO selector is in itself strategyproof with respect to performance misreporting, ignoring the performance of sellers and the task size of queries results in the frequent situation in which large tasks are assigned to slow sellers. For each of the heuristics below, we attempt to avoid this situation.

\subsubsection{Greedy Fastest}

In the greedy fastest selector, we match a query with the fastest available sellers. By doing so, we attempt to maximize the throughput of the system by maximally utilizing the fastest available devices. Interestingly, this heuristic actually leads to decreased utilization of fast devices. By attempting to quickly satisfy all queries, fast sellers quickly leave the computation pool and only slow sellers remain in times of high load.

\subsubsection{SP Fastest/Slowest On-Time \label{sec:Almost-Nash-Algorithm}}

In order to maximally utilize our fast sellers and minimize the probability that a query is fulfilled slower than the query device could have computed the result itself (e.g. maximize the probability of ``on-time'' queries), we propose two heuristics: fastest on-time and slowest-on time.\\
\\
In the fastest on-time heuristic, we match a query with the fastest devices such that both matched devices are fast enough to fulfill the query on-time. If no such sellers are available, we defer matching of the query until the next round. If the performance of a seller required to fulfill a query is faster than all sellers, we match the query with the fastest available sellers regardless of their performance.\\
\\
In the slowest on-time heuristic, we match a query with the slowest devices such that both matched devices are fast enough to fulfill the query on-time. In the way, we still attempt to fulfill queries on-time while keeping faster sellers in the system longer and buffering against spikes of large tasks. In doing so, we sacrifice average wait time in low load situations. However, in the event of high load, the large proportion of available fast sellers results in a net decrease in average wait time as we better avoid the situation in which large tasks are matched to slower-performing devices.\\
\\
In a particular instance of the SP slowest on-time selector, paired with a immediate matcher and hardest-to-fulfill query reorderer, we empirically show that the result eventually demonstrates Nash Equilibrium with high probability, such that for each device the best option is to remain in the system to submit queries.

\section{Approximation of Theoretically Optimal Algorithm \label{sec:Deriving-Theoretically-Optimal}}

In the subsections below, we derive a distributional model under a known query distribution to obtain a model that allows us to match current queries with future sellers. By doing so, we minimize the expected wait time across all queries, taking into account the probability that any given seller will query a product and enter the marketplace.

\subsection{Theoretical Optimal Solution for Unmatched Queries \label{sec:Theoretical-Base}}

Suppose for our optimization metric we choose to maximize the expected gain across all unmatched queries. In deriving a theoretically optimal algorithm, we assume the following:
\begin{enumerate}
\item The query distribution $\pi(\theta_1)$ is known and fixed.
\item The performance of each device is known.
\item Computation time is a deterministic function of device performance and task size, namely $\phi(p, q) = \frac{Tok(q)}{Perf(p)}$ for device $p$ and query $q$.
\item No partial matchings are allowed -- a query must be matched to two sellers before either seller computation can begin.
\item We have an infinite number of products, where each query is of a distinct product.
\end{enumerate}
Then for any matching, we have the following maximization problem:
\begin{gather*}
M^* = \argmax_{M \in \mathcal{M}} E[\sum_{(q, p_1, p_2, k_1, k_2) \in M} (2\phi(Buyer(q), q) - w(q, p_1, p_2, k_1, k_2))]
\end{gather*}
By linearity of expectation, the maximization problem above is equivalent to maximizing the sum of expected gains across all possible matchings. Also, note that $\sum \phi(Buyer(q), q)$ is a constant across all possible matchings, and for optimization purposes we can ignore it. Thus we have:
\begin{gather*}
M^* = \argmin_{M \in \mathcal{M}} \sum_{(q, p_1, p_2, k_1, k_2) \in M} E[w(q, p_1, p_2, k_1, k_2)]
\end{gather*}
Additionally, we can decompose the wait time for a query $w(q, p_1, p_2, k_1, k_2)$ as follows:
\begin{align*}
w(q, p_1, p_2, k_1, k_2) &= \max(\psi(p_1) + \beta(q, p_1, k_1), \psi(p_2) + \beta(q, p_2, k_2)) \text{, where}\\
\\ \beta(q, p, k) &= \phi(p, q) + \tau(p, k)
\end{align*}
Here, we define $\tau(p, k)$ as the recursive function that represents the total time until device $p$ is idle for the $k$th person in line waiting on $p$. Let $q^+$ denote the query ahead of $q$ in line waiting for device $p$, and $\xi_{k}(p)$ denote the time until device $p$ becomes idle starting at the time when $p$ finishes computation for the $k$ queries ahead of $q$ in line. Then:
\begin{align*}
\tau(p, k) =
\begin{cases}
\xi(p) & \text{if } k = 0\\
\xi_{k}(p) + \beta(p, q^+, k-1) & \text{if } k > 0
\end{cases}
\end{align*}
From this, we obtain the optimal matching $M^*$:
\begin{align*}
M^* &= \argmin_{M \in \mathcal{M}} \sum_{(q, p_1, p_2, k_1, k_2) \in M} E[\max(\psi(p_1) + \beta(q, p_1, k_1), \psi(p_2) + \beta(q, p_2, k_2))]\\
&= \argmin_{M \in \mathcal{M}} \sum_{(q, p_1, p_2, k_1, k_2) \in M} E[\max(\psi(p_1) + \phi(p_1, q) + \tau(p_1, k), \psi(p_2) + \phi(p_2, q) + \tau(p_2, k)]
\end{align*}\\

\subsection{Approximation for Exponential Query Distribution \label{sec:Exponential-Approximation}}

Suppose we assume that queries frequencies follow an exponential distribution with parameter $\alpha$ such that $\pi(x;\theta_1) = f(x;\alpha) = \alpha e^{-\alpha x}$. In order to make the above problem tractable, we approximate $\tau(p, k)$ with $\xi(p)$ and break ordering ties by assigning each incoming query a random priority and ordering conflicting matchings from highest to lowest priority. Then, given the tasks that can be started after we obtain our optimal matching, we iteratively recompute the optimal matching on our remaining queries until convergence. By using the expectation possible future query times, we allow a query to match with future devices and wait for them to appear in the marketplace.\\
\\
We now utilize our exponential distribution over query frequency to formulate our matching algorithm. Substituting in our approximation for $\tau$, we obtain:
\begin{align*}
M^* &= \argmin_{M \in \mathcal{M}} \sum_{(q, p_1, p_2, k_1, k_2) \in M} E[\max(\psi(p_1) + \phi(p_1, q) + \xi(p_1), \psi(p_2) + \phi(p_2, q) + \xi(p_2)]
\end{align*}
We then have three cases for our expectation. In the first, suppose both $p_1$ and $p_2$ are idle. Then $\xi(p_1) = \xi(p_2) = 0$, and our expectation is simply the maximum of two constant valued expressions.\\
\\
In the second case, suppose one of $\xi(p_1)$ or $\xi(p_2)$ is 0. We then have the expected maximum over a constant-valued expression and a shifted exponential distribution. As an example, suppose that $\psi(p_1) + \phi(p_1, q) > \psi(p_2) + \phi(p_2, q)$ and $\xi(p_1) = 0$. Then:\\
\begin{align*}
&E_{f(\lambda;\alpha)}[\max(\psi(p_1) + \phi(p_1, q) + \xi(p_1), \psi(p_2) + \phi(p_2, q) + \lambda)]\\
=&\psi(p_1) + \phi(p_1, q) + E_{f(\lambda;\alpha)}[\max(0, \psi(p_2) + \phi(p_2, q) - \psi(p_1) - \phi(p_1, q) + \lambda)]\\
=& \psi(p_1) + \phi(p_1, q) + \alpha e^{-\alpha (\psi(p_1) + \phi(p_1, q) -\psi(p_2) - \phi(p_2, q))}\\
\end{align*}
In the final case, suppose neither device is currently idle. Furthermore, without loss of generality let $\psi(p_1) + \phi(p_1, q) \geq \psi(p_2) + \phi(p_2, q)$. Then we have:
\begin{align*}
&E_{f(\lambda_1, \lambda_2;\alpha)}[\max(\psi(p_1) + \phi(p_1, q) + \lambda_1, \psi(p_2) + \phi(p_2, q) + \lambda_2)]\\
=&\int_{\lambda = 0}^{\infty} \alpha \lambda (e^{-\alpha (\lambda - \psi(p_1) - \phi(p_1, q))} - 2 e^{-\alpha (2 \lambda - \psi(p_1) - \phi(p_1, q) - \psi(p_2) - \phi(p_2, q))} + e^{-\alpha(\lambda - \psi(p_2) - \phi(p_2, q))}) d\lambda\\
=& \alpha^{-1} + \psi(p_2) + \phi(p_2, q) + \frac{1}{2} e^{-\alpha(\psi(p_1) + \phi(p_1, q) - \psi(p_2) - \phi(p_2, q))}
\end{align*}
From these expressions, we are able to compute the approximate optimal solution by selecting the pairing $(p_i, p_j)$ for each unfilled query $q$ that minimizes the expected waiting time, breaking ties using random priorities as described above.

\section{Experimental Results \label{sec:Experiment-Results}}

\paragraph*{\textmd{We implemented several combinations of heuristic match schedulers, reorderers, 
and selectors for comparison. While we did successfully implement the distributionally based algorithm in Section \ref{sec:Deriving-Theoretically-Optimal}, the poor scalability of the algorithm with a larger number of devices precludes it from useful comparison. Each algorithm was evaluated in a simulated environments using the following configurations:
query frequency distribution $\pi(\theta_{1}):=Uniform(0,1\,hour)$,
device performance distribution $\sigma(\theta_{2}):=Uniform(1,100)$
(tokens per millisecond), task size distribution $\omega(\theta_{3}):=Uniform(10000,5000000)$
(tokens). The number of devices $|P|=100,000$, the number of servers
$\mu=2$, and the performance of the servers are fixed at $1,000$
tokens per millisecond. For simplicity, we assume an infinite
number of products such that there is a one-to-one mapping between buyers
and queries. For all algorithms we disable partial matching (allow
matching query with one seller first and start computation), except
for InstantSPReversedImprovedMatcher and InstantSPImprovedMatcher.
For each experiment simulates a full day of simulation and provides the following: 
moving average wait time (in groups of 100 elements) with
its histogram, raw wait time with histogram, time saved v.s device
performance scatterplot, idle seller performance histogram, slow
query ratio per phone histogram, and match time. We also compute a
series of summary statistics, including metrics mentioned in Section \ref{sec:Evaluation-Metrics}
to compare multiple algorithms. The results are shown
in Table \ref{tab:Performance-metrics-comparison} and figures in
Appendix \ref{sec:Appendix} (a missing subplot indicates lack of
sufficient amount of data).}}

\paragraph*{\textmd{From the results, we can see }InstantSPReversedImproved \textmd{is
the most likely algorithm to be in a state of Nash Equilibrium (with respect to expected benefit of 
remaining in the system) as it has negligible slow query
ratio, minimal }MaxNetLoss\textmd{, }MaxTimeWasted\textmd{, and }AverageTimeWasted\textmd{
in absolute values. The strategy of matching devices with the slowest
sellers that are just good enough is proven to be effective, as }InstantSPImproved\textmd{
is showing consistently better performance in }MaxNetLoss\textmd{,
}MaxTimeWasted\textmd{, and }AverageTimeWasted. \textmd{Partial matching
is also proven to be effective by the comparison between }InstantSPImproved
\textmd{and }InstantSP.\textmd{ In addition, scheduled matchers 
appear to be weaker than instant matchers in the presence of good heuristics 
as can be seen from
the comparison between }InstantSP \textmd{and }ScheduledSP\textmd{.
Between those matchers optimizing for wait time, }ScheduledMinVar\textmd{
stands out by a large margin. Probablistic matchers are close in
average waiting time, but the actual computation time is generally much longer than 
with non-probabilistic matchers. All non-SP matchers give unsatisfactory results in
terms of slow query ratio, time wasted, and maximum net loss. See
Figures in Appendix \ref{sec:Appendix} for details.}}

\begin{table}

\begin{centering}
\begin{tabular}{|c|c|c|c|c|}
\hline 
 & AvgWait (s) & MaxTimeSaved & AvgTimeSaved & SlowRatio\tabularnewline
\hline 
 & MaxWait(s) & MaxTimeWasted & AvgTimeWasted & MaxNetLoss\tabularnewline
\hline 
\hline 
\multirow{2}{*}{InstantSPReversedImproved} & 114.62 & 9000 & 88.986 & 1.718e-5\tabularnewline
\cline{2-5} 
 & 4855.93 & -0.401 & -0.245 & 0.0\tabularnewline
\hline 
\multirow{2}{*}{InstantSPImproved} & 105.068 & 9111.65 & 105.611 & 2.089e-5\tabularnewline
\cline{2-5} 
 & 4818.518 & -8.156 & -1.631 & 0.0\tabularnewline
\hline 
\multirow{2}{*}{InstantSP} & 105.525 & 9599 & 109.314 & 3.339e-4\tabularnewline
\cline{2-5} 
 & 3290.716 & -28.714 & -3.959 & -693\tabularnewline
\hline 
\multirow{2}{*}{ScheduledSP} & 113.56 & 9178.83 & 89.507 & 0.0011\tabularnewline
\cline{2-5} 
 & 26648.87 & -26656.798 & -8276 & -34598\tabularnewline
\hline 
\multirow{2}{*}{ScheduledMinVar} & 53.76 & 9429 & 252.329 & 0.258\tabularnewline
\cline{2-5} 
 & 81864.42 & -81794 & -27.566 & -81807\tabularnewline
\hline 
\multirow{2}{*}{SelectiveProbablisticScheduledGreedy} & 61.635 & 9795 & 206 & 0.057\tabularnewline
\cline{2-5} 
 & 14945 & -14695 & -539 & -14695\tabularnewline
\hline 
\multirow{2}{*}{InstantFIFO} & 109.865 & 9679 & 223 & 0.269\tabularnewline
\cline{2-5} 
 & 4823 & -4537 & -153 & -11239\tabularnewline
\hline 
\multirow{2}{*}{InstantGreedy} & 73 & 9571 & 203 & 0.163\tabularnewline
\cline{2-5} 
 & 1376 & -1101 & -79 & -4346\tabularnewline
\hline 
\end{tabular}\protect\caption{\label{tab:Performance-metrics-comparison} Performance metrics comparison
between algorithms. The first three columns are taken from the statistics
over the last \textasciitilde{}250000 queries. The last column is
taken from the cumulative statistics of the entire run. All units
are in seconds or fractions.}

\par\end{centering}

\end{table}

\section{Conclusions and Future Work \label{sec:Conclusions-and-Future}}

To conclude, we have formulated a new problem space that extends to general situations in which a user wishes to repeatedly perform low-bandwidth, computationally-intensive tasks. While the Chital system was designed with LDA in mind, it is easy to see how the system could prove useful in any number of machine learning settings that require multiple restarts to mitigate the risk of bad local optima (e.g. neural networks, clustering). We defined several evaluation metrics of matcher performance and described heuristics that optimize for these evaluation metrics while attempting to obtain Nash Equilibrium and enforce strategyproofness of performance reporting and querying. Additionally, we showed how assuming a known distribution over incoming queries can theoretically improve matcher performance by allowing the matcher to perform probabilistically sound pairings between current unfilled queries and future sellers.\\
\\
For future research, it would be interesting to explore the modeling of our optimization problem using physical systems. For example, we may choose to represent the matching problem as an electrical circuit in which each resistor represents the amount of time required to complete a task. Additional work on improving the speed of the algorithm in \ref{sec:Exponential-Approximation} may prove useful in scaling up the algorithm to larger scale, realtime systems. We also wish to formally prove the state of Nash equilibrium in our SP matchers. While the immediate matching SP slowest on-time matcher appeared to \textit{eventually} reach a state in which all users had net positive gain from the system under our tested situations, a formal proof on the conditions under which this state will be achieved would be useful in better characterizing the underlying properties of the matcher across a broader class of situations.

\newpage{}
\fancyhf{}

\begin{centering}
\section*{Appendix \label{sec:Appendix}}
\end{centering}

\begin{figure}[!hbt]
\begin{centering}
\includegraphics[width=1\textwidth]{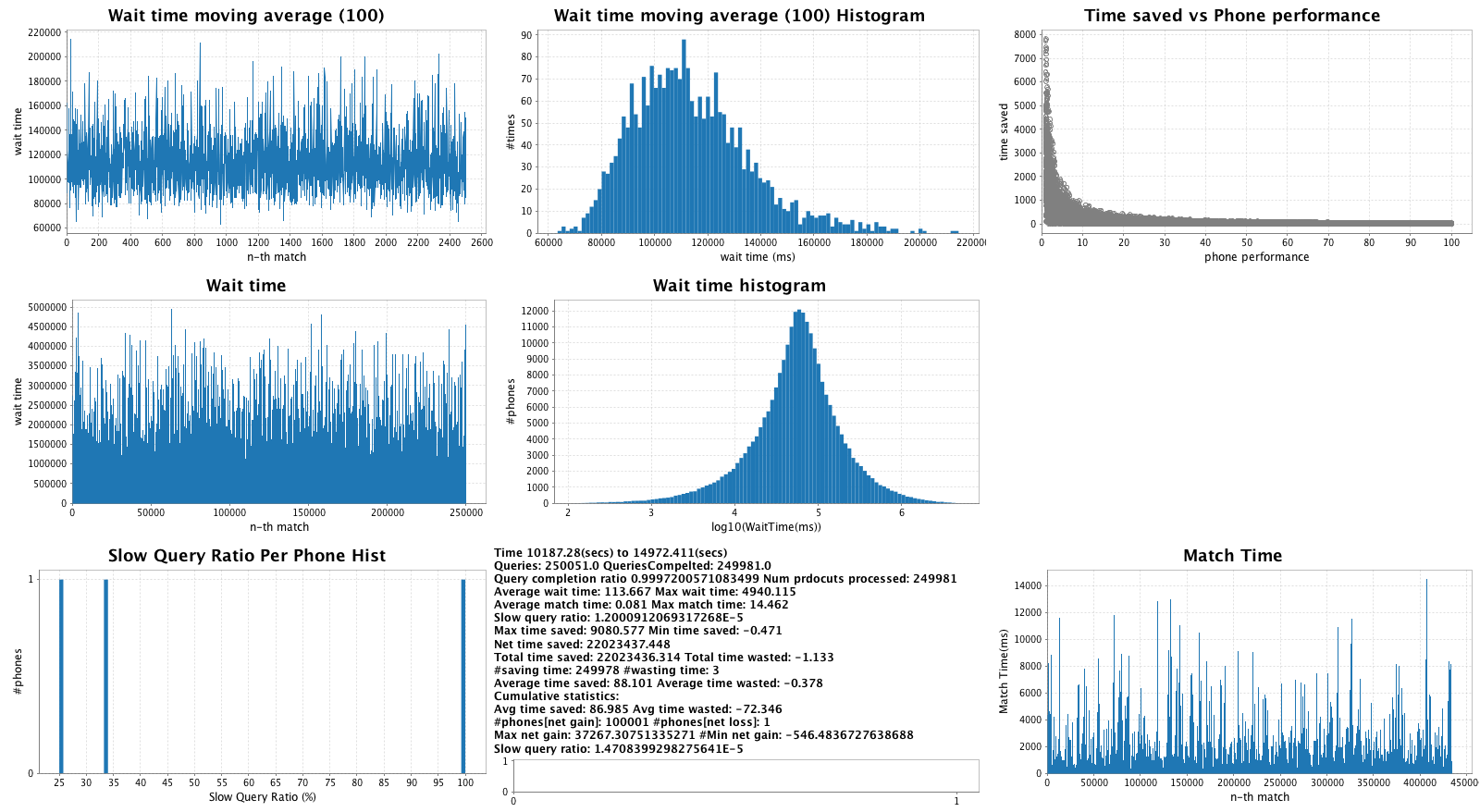}
\par\end{centering}

\protect\caption{InstantSPReversedImproved, Beginning of day }
\end{figure}

\begin{figure}[!hbt]
\begin{centering}
\includegraphics[width=1\textwidth]{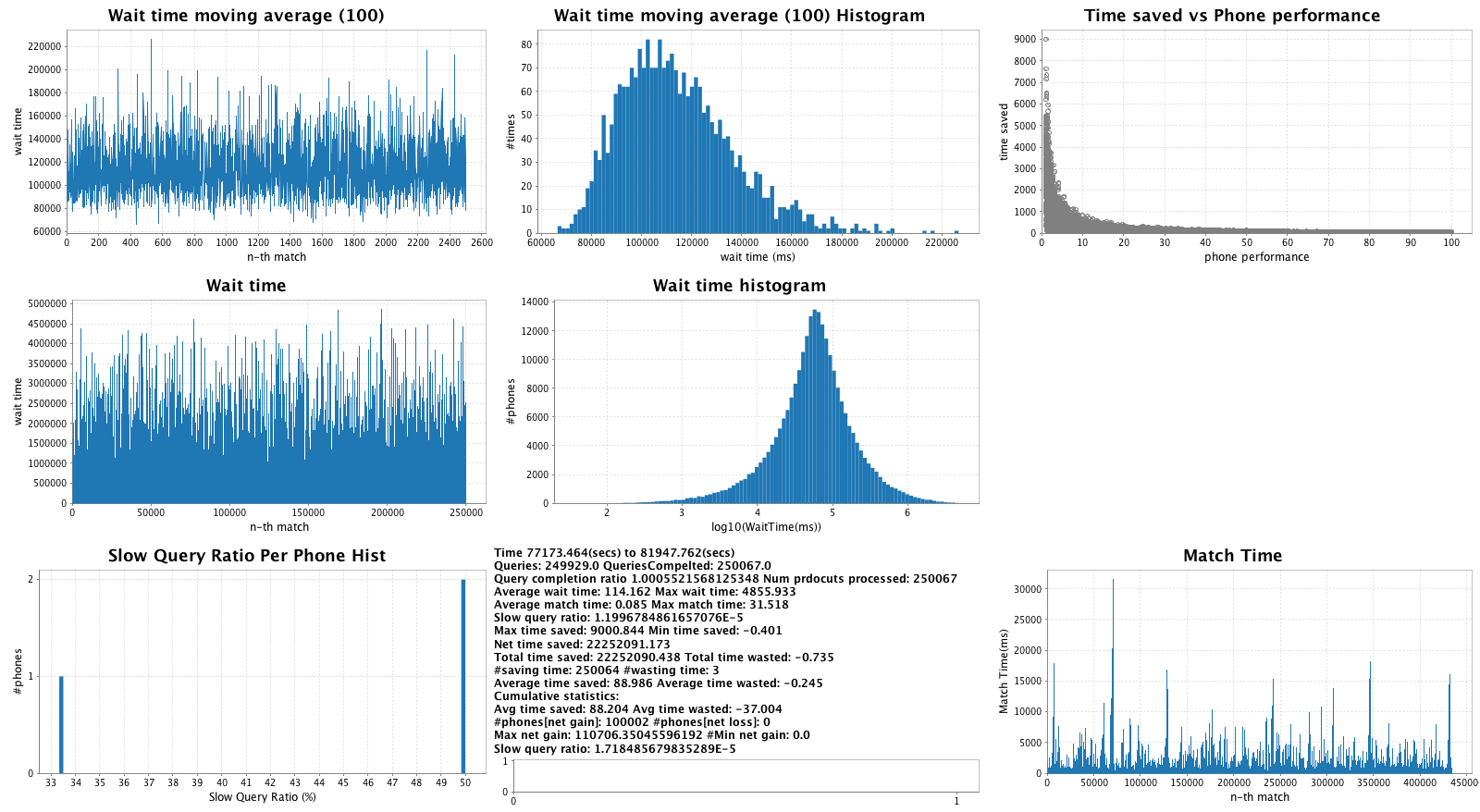}
\par\end{centering}

\protect\caption{InstantSPReversedImproved, End of Day }
\end{figure}

\begin{figure}[!hbt]
\begin{centering}
\includegraphics[width=1\textwidth]{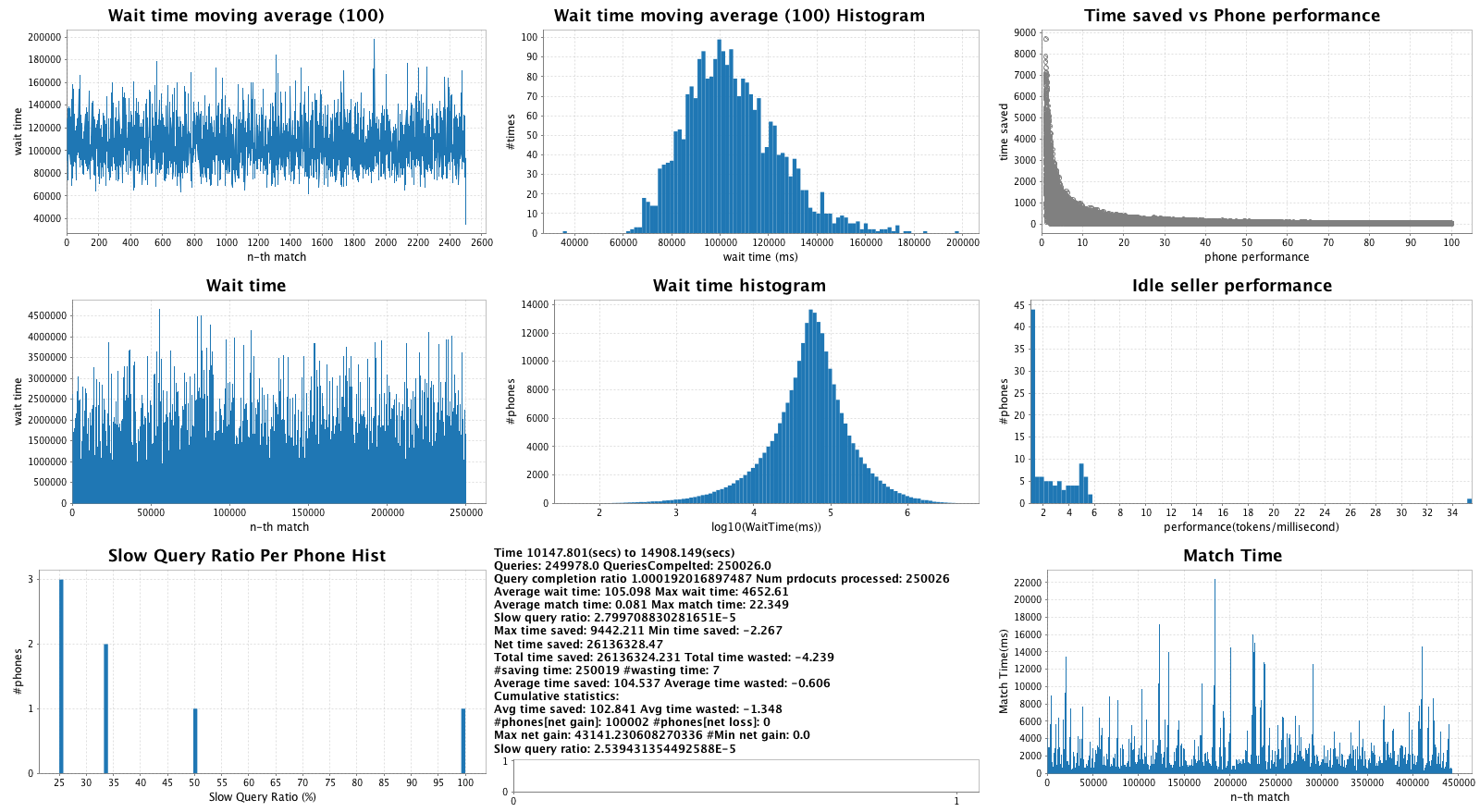}
\par\end{centering}

\protect\caption{InstantSPImproved, Beginning of Day }
\end{figure}

\begin{figure}[!hbt]
\begin{centering}
\includegraphics[width=1\textwidth]{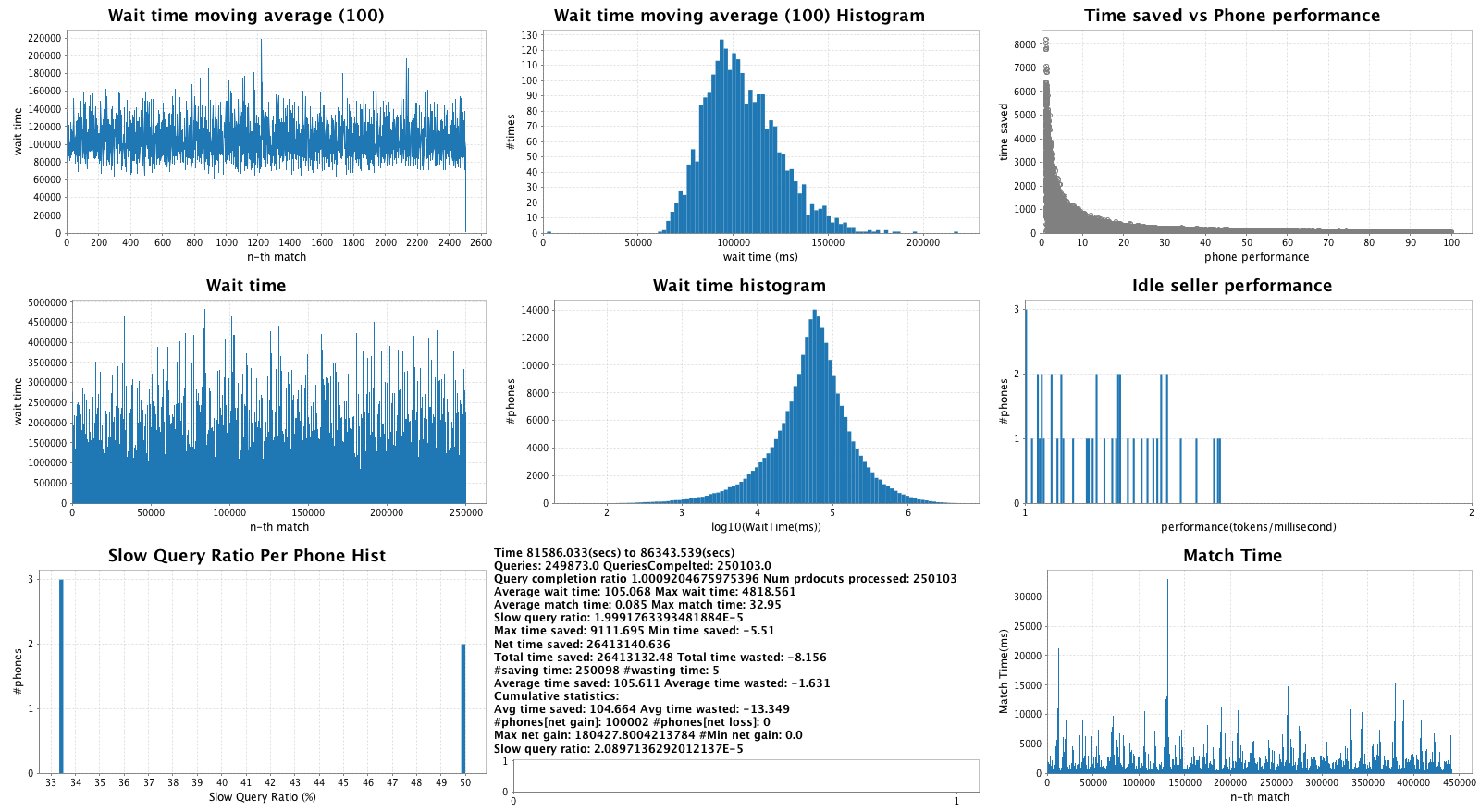}
\par\end{centering}

\protect\caption{InstantSPImproved, End of Day }
\end{figure}

\begin{figure}[!hbt]
\begin{centering}
\includegraphics[width=1\textwidth]{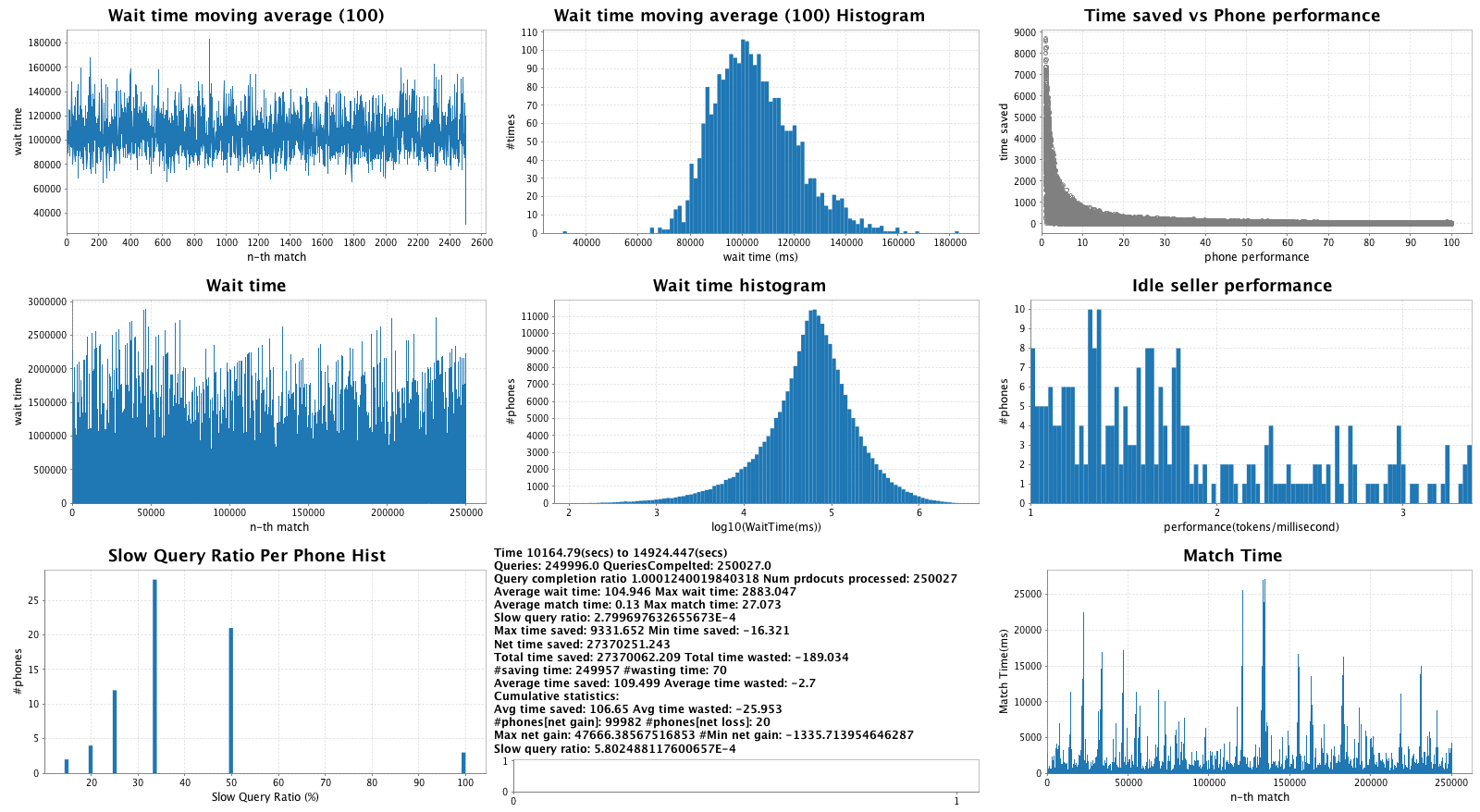}
\par\end{centering}

\protect\caption{InstantSP, Beginning of Day }
\end{figure}

\begin{figure}[!hbt]
\begin{centering}
\includegraphics[width=1\textwidth]{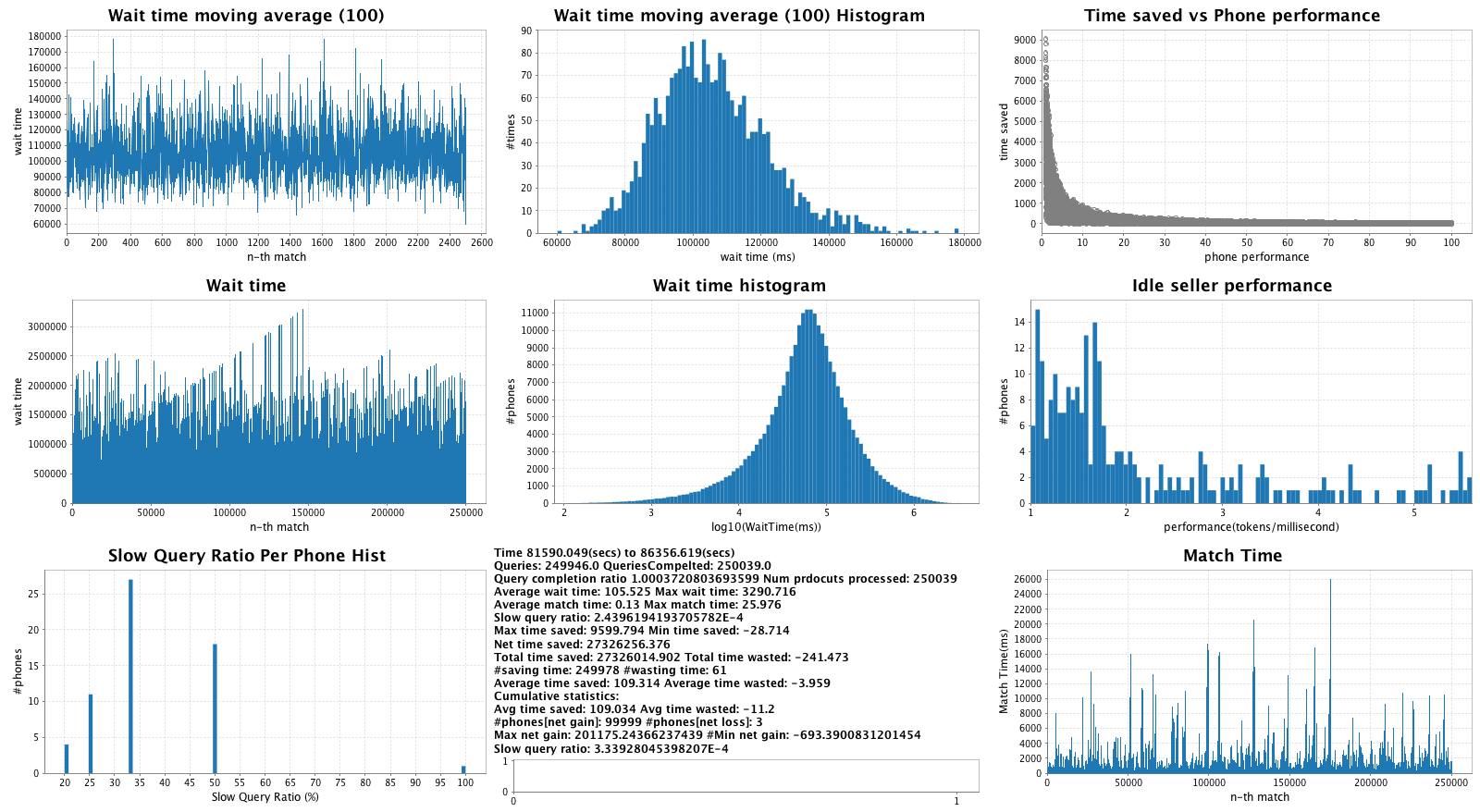}
\par\end{centering}

\protect\caption{InstantSP, End of Day }
\end{figure}

\begin{figure}[!hbt]
\begin{centering}
\includegraphics[width=1\textwidth]{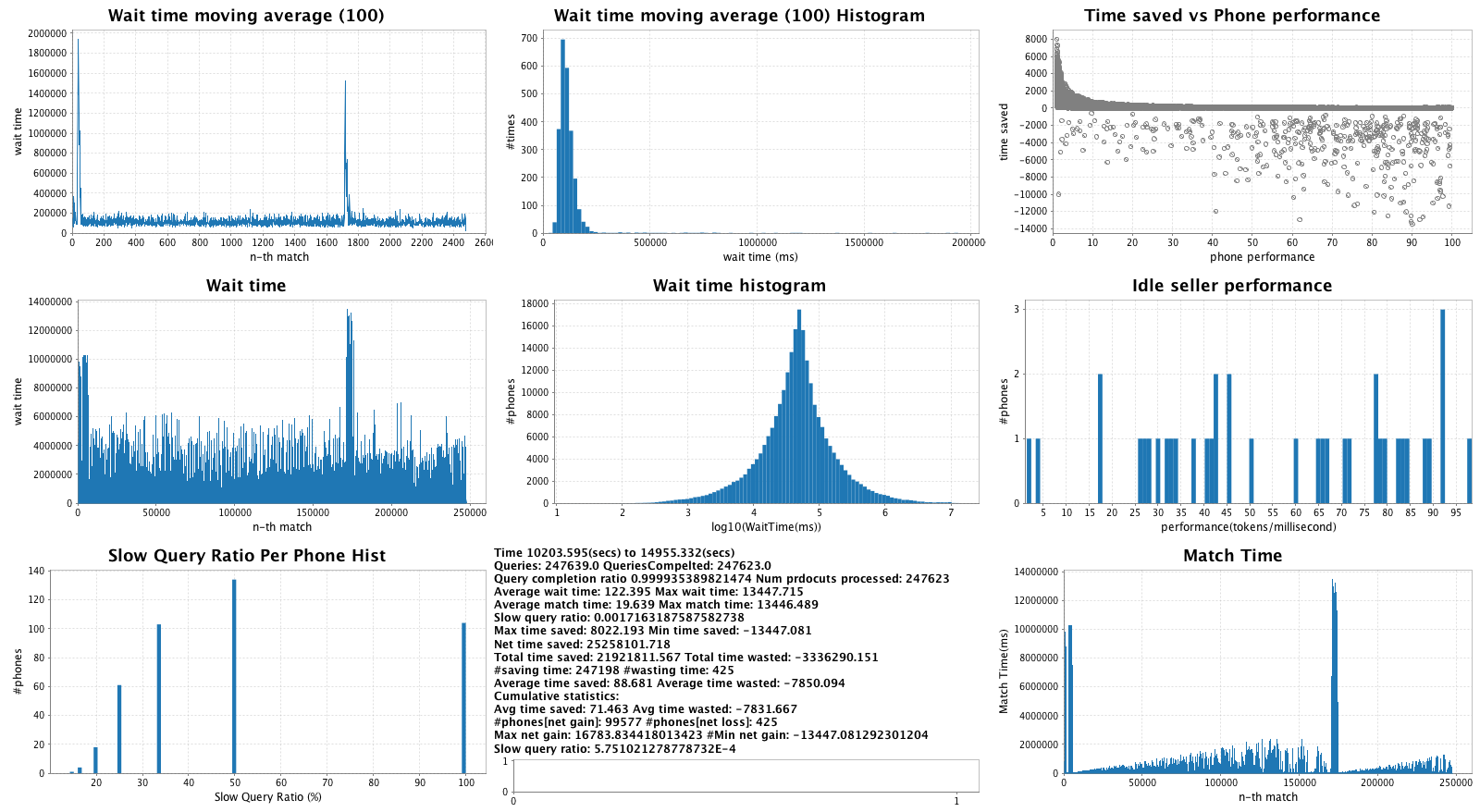}
\par\end{centering}

\protect\caption{ScheduledSP, Beginning of Day }
\end{figure}

\begin{figure}[!hbt]
\begin{centering}
\includegraphics[width=1\textwidth]{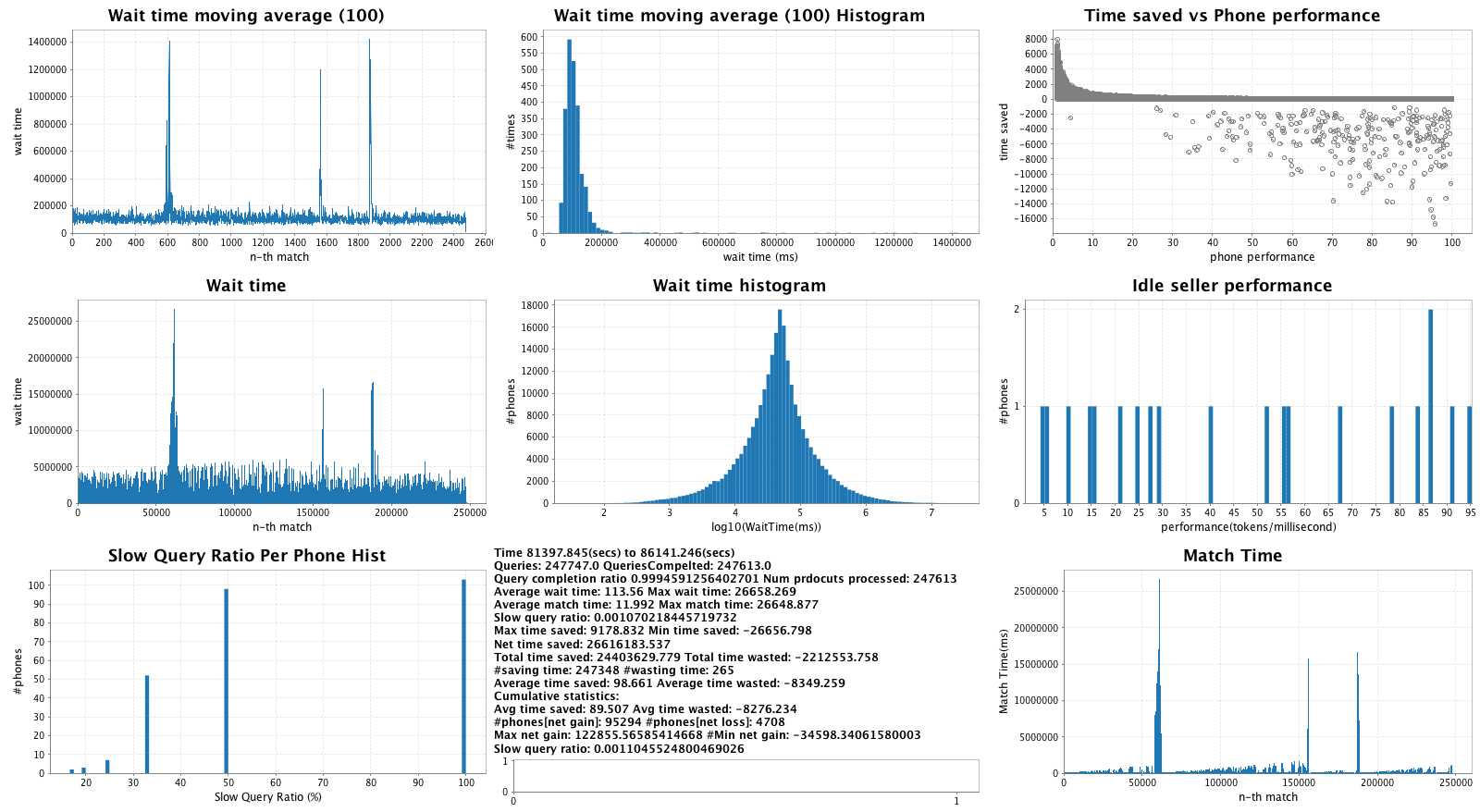}
\par\end{centering}

\protect\caption{ScheduledSP, End of Day }
\end{figure}

\begin{figure}[!hbt]
\begin{centering}
\includegraphics[width=1\textwidth]{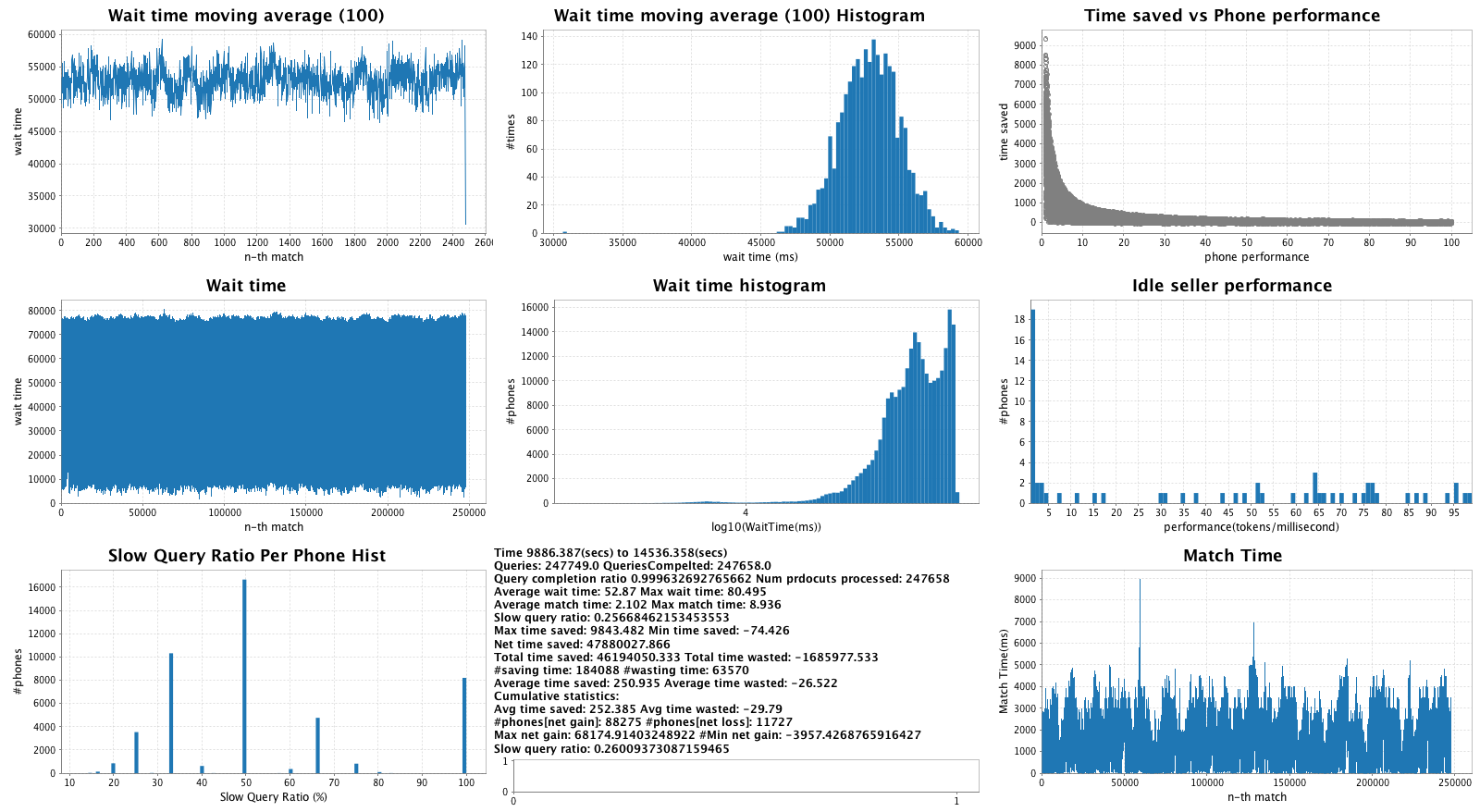}
\par\end{centering}

\protect\caption{SelectiveScheduledMinVarianceMatcher, Beginning of Day }
\end{figure}

\begin{figure}[!hbt]
\begin{centering}
\includegraphics[width=1\textwidth]{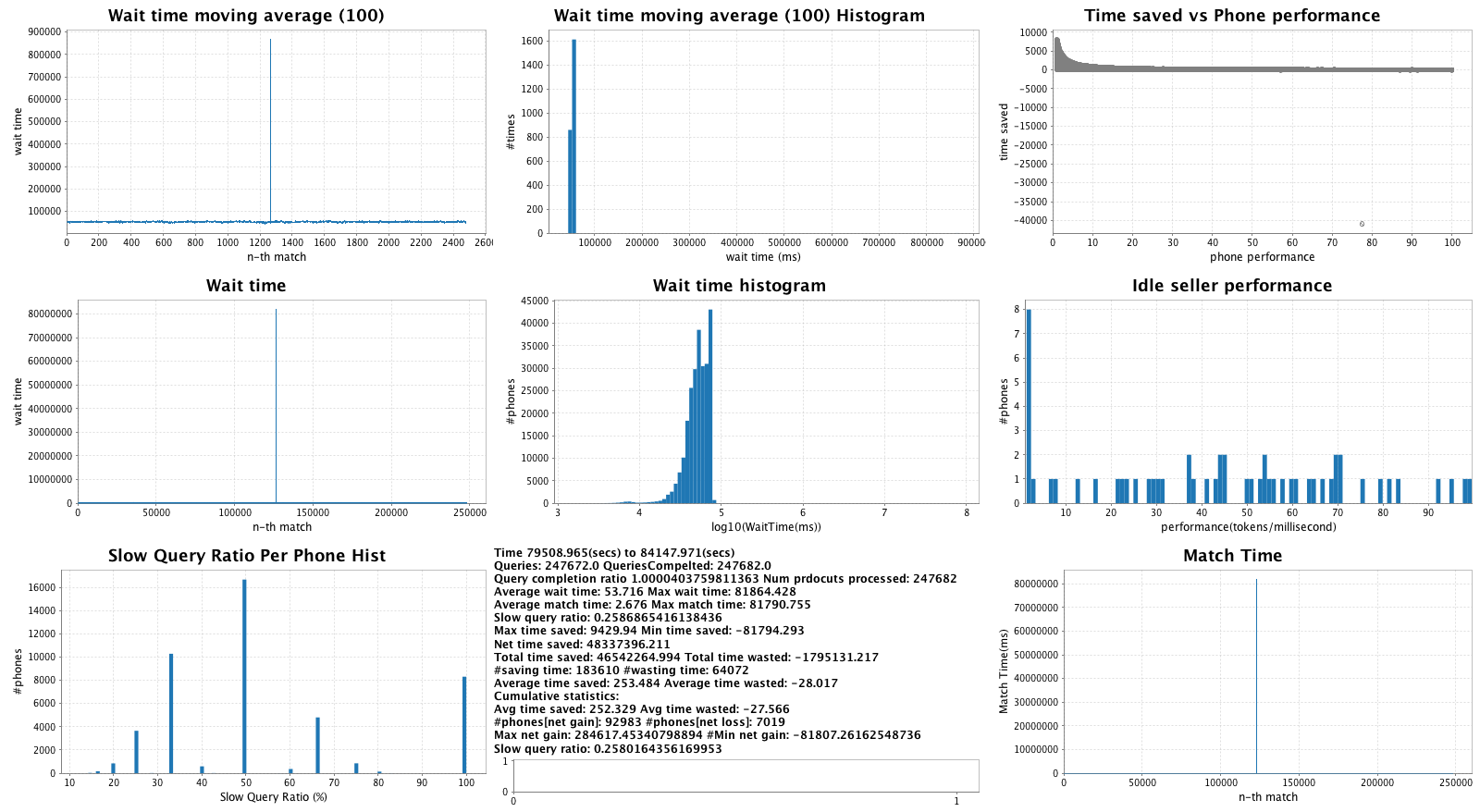}
\par\end{centering}

\protect\caption{SelectiveScheduledMinVarianceMatcher, End of Day}
\end{figure}

\begin{figure}[!hbt]
\begin{centering}
\includegraphics[width=1\textwidth]{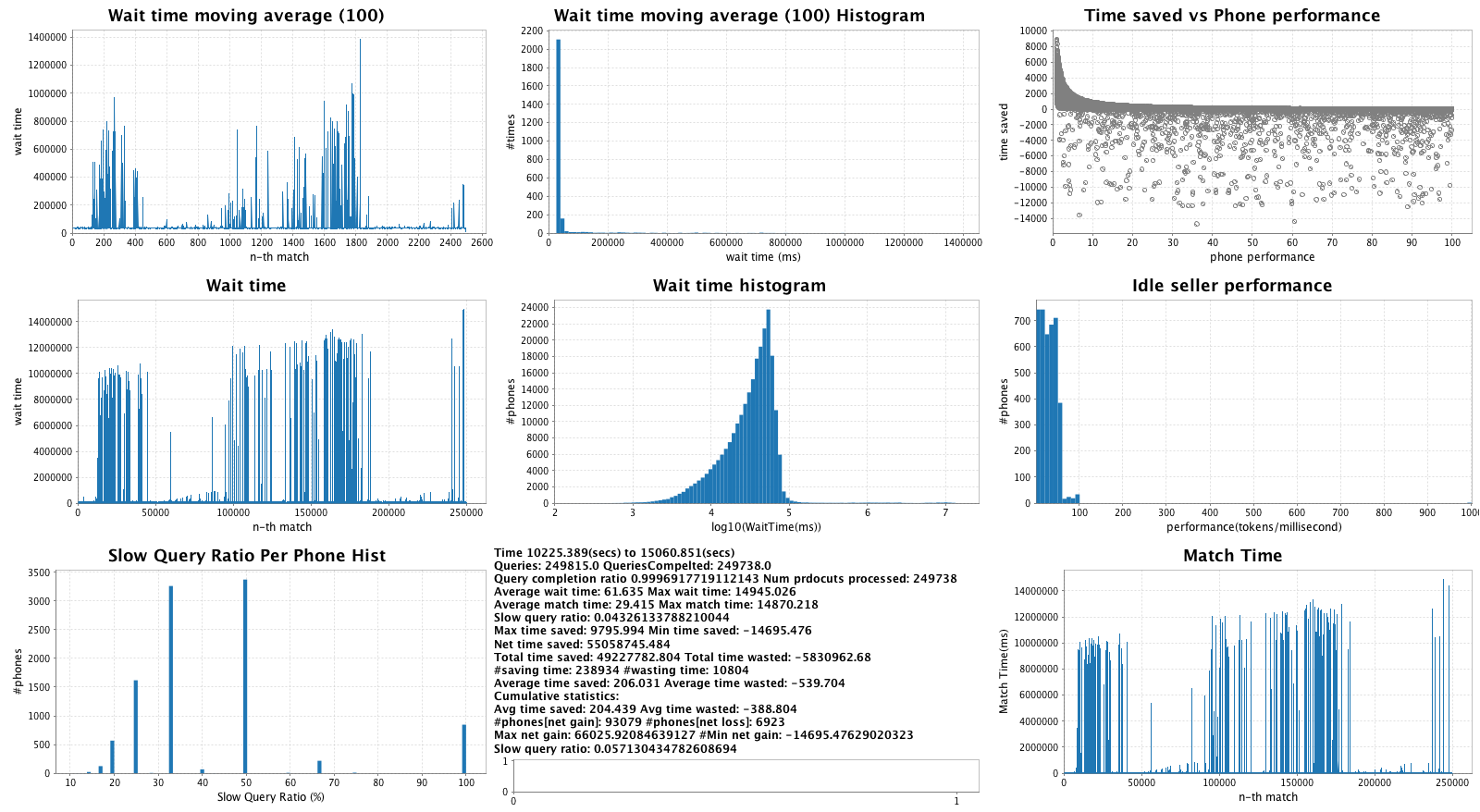}
\par\end{centering}

\protect\caption{SelectiveProbablisticScheduledGreedyMatcher, Beginning of Day}
\end{figure}

\begin{figure}[!hbt]
\begin{centering}
\includegraphics[width=1\textwidth]{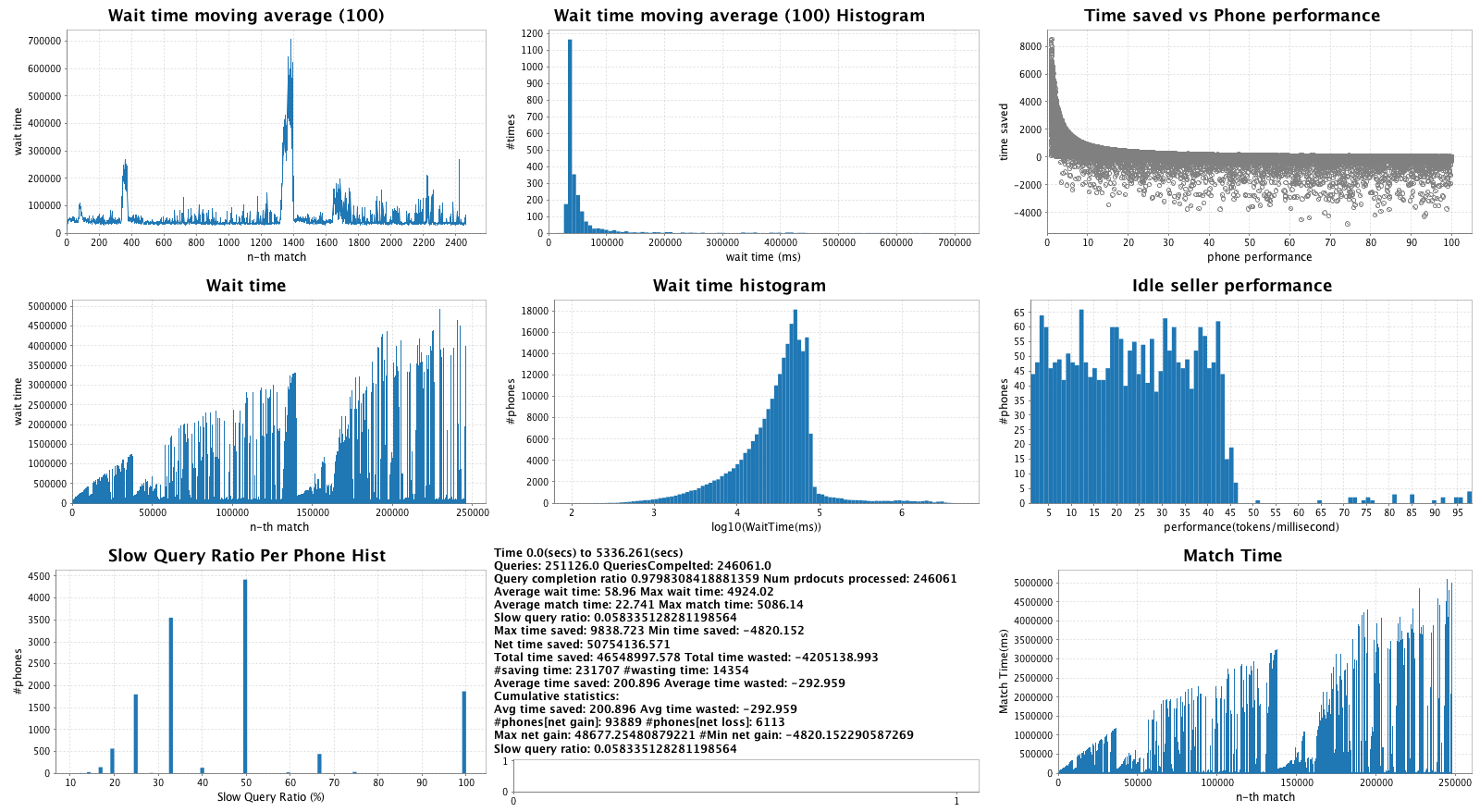}
\par\end{centering}

\protect\caption{SelectiveScheduledGreedyMatcher, Beginning of Day (earlier
interval compared to others due to slow running speed of this algorithm)}
\end{figure}

\begin{figure}[!hbt]
\begin{centering}
\includegraphics[width=1\textwidth]{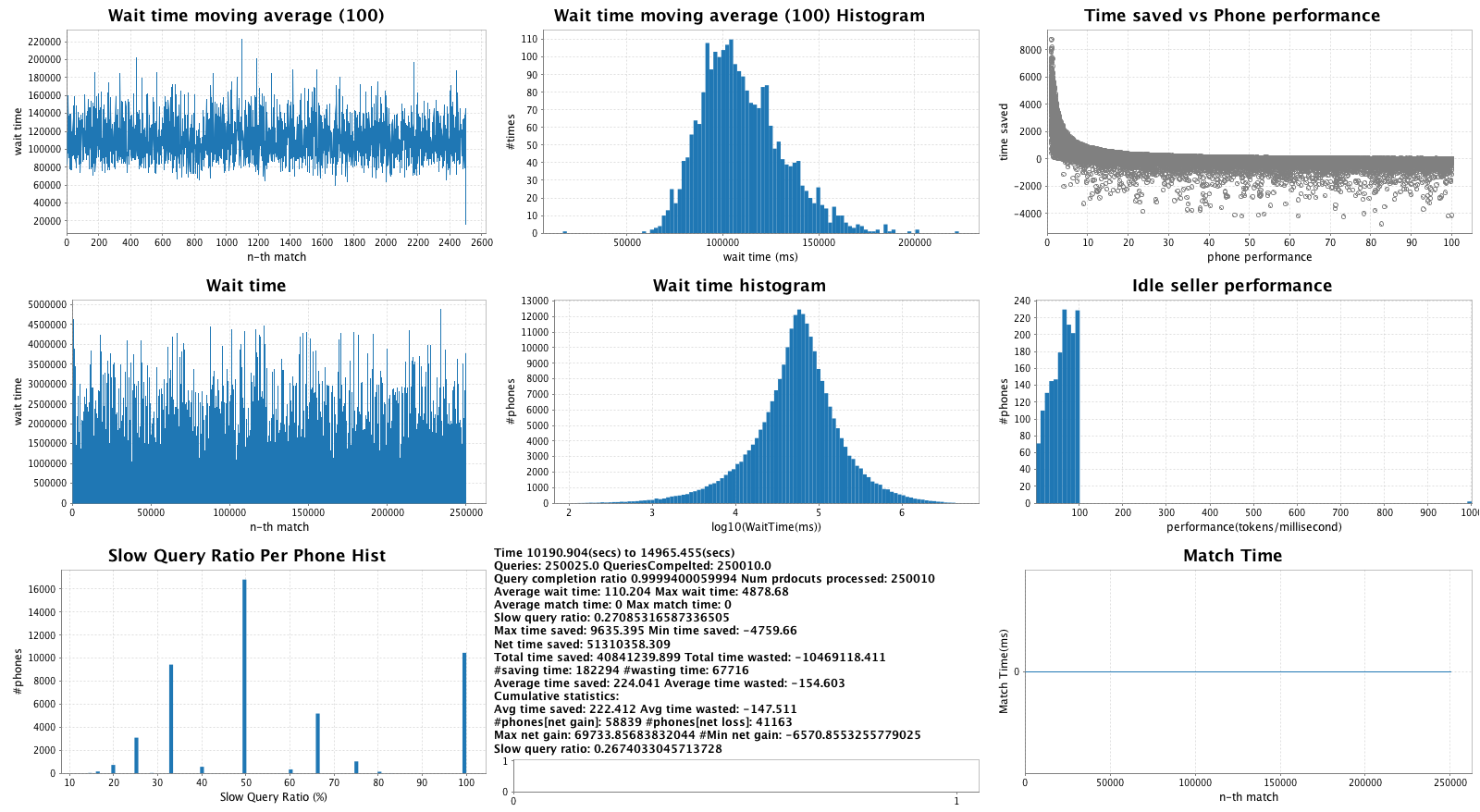}
\par\end{centering}

\protect\caption{InstantFIFOMatcher, Beginning of Day}
\end{figure}

\begin{figure}[!hbt]
\begin{centering}
\includegraphics[width=1\textwidth]{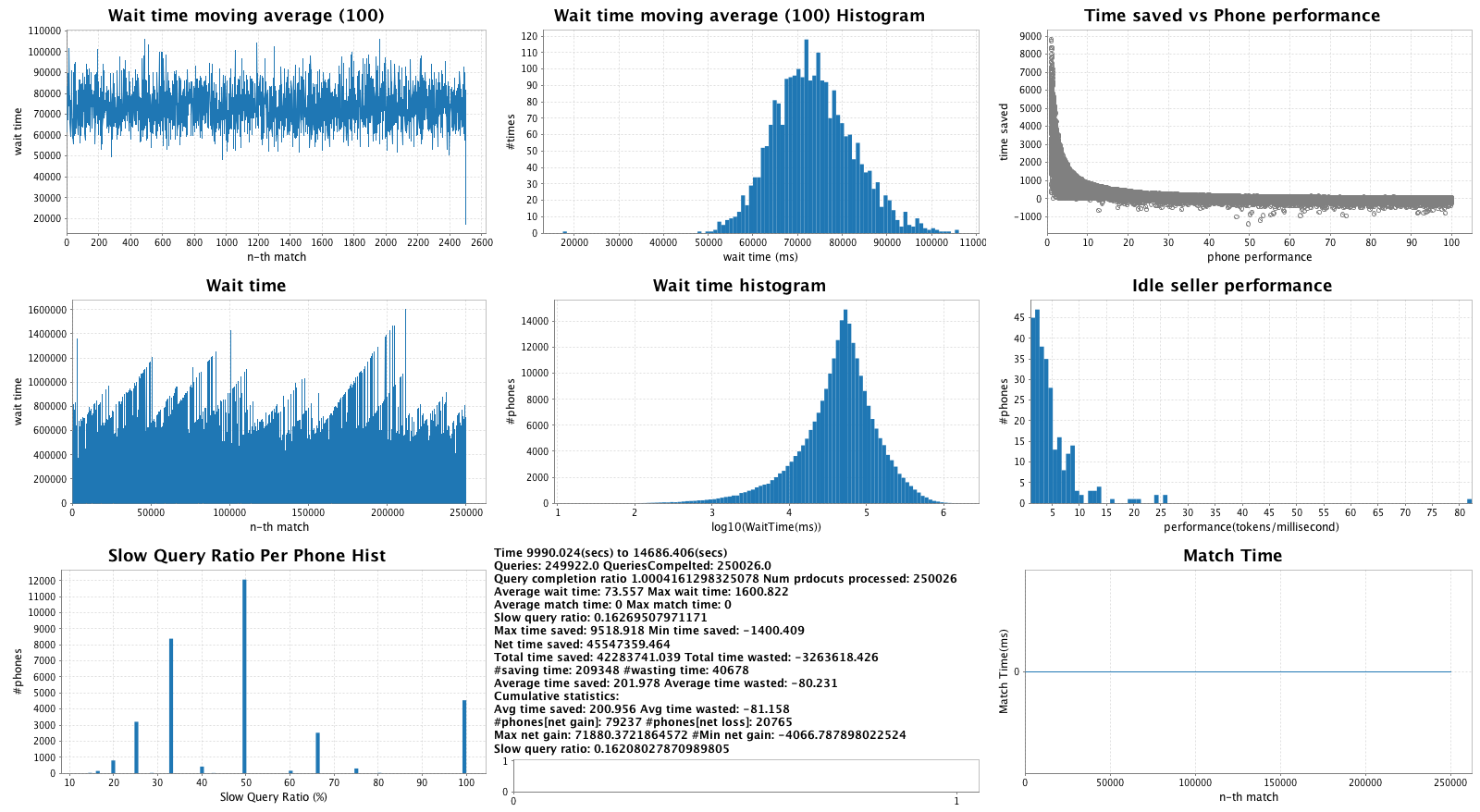}
\par\end{centering}

\protect\caption{InstantGreedy, Beginning of Day}
\end{figure}

\begin{figure}[!hbt]
\begin{centering}
\includegraphics[width=1\textwidth]{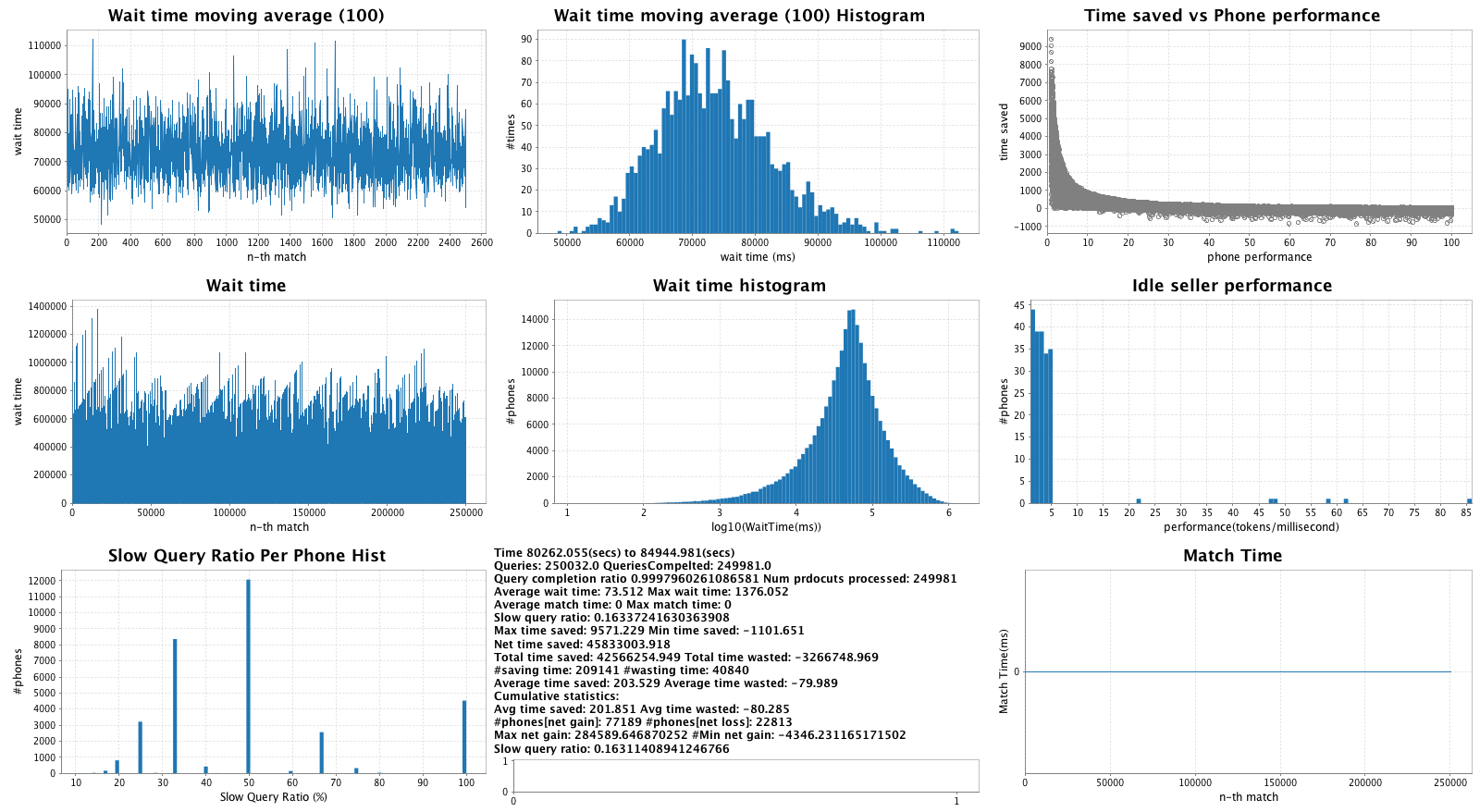}
\par\end{centering}

\protect\caption{InstantGreedy, End of Day}
\end{figure}


\begin{thebibliography}{10}
\bibitem{LietalANU2012} Aaron Q Li, \newblock Multi-GPU Distributed
Parallel Bayesian Differential Topic Model. \newblock Australian
National University (Thesis)

\bibitem{LietalWSDM2015} Aaron Q Li, Amr Ahmed, Mu Li, Vanja Josifovski,
Alexander J Smola, \newblock High Performance Latent Variable Models.
\newblock WWW 2015 (Submitted).

\bibitem{LietalKDD2014} Aaron Q Li, Amr Ahmed, Sujith Ravi, Alexander
J Smola, \newblock Reducing Sampling Complexity of Topic Models.
\newblock ACM SIGKDD 2014.

\bibitem{715Project} Aaron Q Li, Joseph W Robinson, Yuntian Deng,
Kublai Jing, \newblock Creating Scalable and Interactive Web Applications
Using High Performance Latent Variable Models. \newblock CMU 10-715
Final Project.

\bibitem{MALLET} Andrew Kachites McCallum, \newblock MALLET: A Machine
Learning for Language Toolkit. \newblock http://mallet.cs.umass.edu.
2002.

\bibitem{OriginalLDA2003} David M Blei, Andrew Y Ng, Michael I Jordan,
\newblock Latent Direchlet Allocation. \newblock Journal of Machine
Learning Research 3.

\bibitem{LietalNIPS2013} Mu Li, Li Zhou, Zichao Yang, Aaron Q Li,
Fei Xia, David G Anderson, Alexander J Smola, \newblock Parameter
Server for Distributed Machine Learning. \newblock NIPS Big Learning
Workshop 2013.

\bibitem{amazondata} Jure Leskovec and Andrej Krevl, \newblock SNAP
Large Network Dataset Collection 2014. \newblock http://snap.stanford.edu/data/web-Amazon-links.html.

\bibitem{Mehta2013} Aranyak Mehta, \newblock Online Matching and
Ad Allocation, Foundations and Trends in Theoretical Computer Science
2013.

\bibitem{KohtRegev2003}Subhash Khot and Oded Regev, Vertex Cover
Might be Hard to Approximate to within $2-\epsilon$, IEEE Conference
on Computational Complexity, 2003.

\bibitem{YajunWangChuwaiWong2013}Yajun Wang and Sam Chiu-wai Wong,
Online Vertex Cover and Matching: Beating the Greedy Algorithm, 2013
\end{thebibliography}
\end{document}